\documentclass[12pt]{article}
\usepackage{graphicx}
\usepackage{epstopdf}
\usepackage{amsmath,amssymb,amsfonts,graphicx,esint}
\begin{document}
\thispagestyle{empty}
\renewcommand{\refname}{References}

\title{\bf Path integral formalism for finite-temperature field theory and generation of chiral currents} 

\author{Yurii A. Sitenko$^{1,2}$}

\date{}

\maketitle
\begin{center}
$^{1}$ Donostia International Physics Center,\\
4 Paseo Manuel de Lardizabal,\\
20018 Donostia-San Sebastián,\\ 
Gipuzkoa, Spain\\
$^{2}$ Bogolyubov Institute for Theoretical Physics,\\
National Academy of Sciences of Ukraine,\\
14-b Metrologichna Street, 03143 Kyiv,\\ Ukraine\\

\end{center}

\begin{abstract}
With the use of the path integral formalism for finite-temperature field theory, I find the persisting vector and axial currents that are generated in quantum chiral fermionic systems. The role of the explicit violation of chiral symmetry by fermion mass is elucidated. For the case of quantum fermionic systems in the backround of an arbitrary smooth magnetic field, I show that the chiral magnetic effect is substantiated on equally the same footing as the chiral separation effect is. Both effects remain unaltered if chiral symmetry is violated by mass.
\end{abstract}

%PACS: 11.10.Wx, 03.70.+k, 71.70.Di, %73.23.Ra, 12.39.Ba, 25.75.Ld

\bigskip

\begin{center}
\end{center}

\bigskip

\section{Introduction}
\setcounter{equation}{0}

Properties of spin-$1/2$ charged particles in the background of a magnetic field were studied comprehensively for almost nine decades after the seminal works of L.D.Landau on diamagnetism and paramagnetism of metals, see \cite{Lan}. Even so two remarkable effects were theoretically discovered relatively recently: the chemical potential generates the chiral imbalance encoded by the axial current along the magnetic field \cite{Met}, while the vector current along the magnetic field is generated by the chiral imbalance encoded by the chiral chemical potential \cite{Fuk}. The first one bears the name of the chiral separation effect, while the second one is known as the chiral magnetic effect. These effects are widely discussed in physics community, with implications especially of the chiral magnetic effect for various areas of contemporary physics, ranging from high energy physics (heavy-ion collisions \cite{Khar}), astrophysics (neutron stars and magnetars \cite{Tur}), cosmology (the early universe \cite{Tash}), and condensed matter physics (novel materials known as the Dirac and Weyl semimetals \cite{Arm}).

It should be noted that the results of \cite{Met} and \cite{Fuk} were obtained for the case of unbounded (infinite) space; they are independent of temperature, being given by rather brief formulas that transform from one another under simultaneous interchange of the axial current with the vector one and of the conventional chemical potential with the chiral chemical one. If space is bounded with a plausible boundary condition imposed, then the fate of the chiral effects is different: the chiral magnetic effect disappears, whereas the chiral separation effect stays on, becoming dependent on the boundary condition, as well as on temperature \cite{Si16a,Si16b,Si18}, see also \cite{Gor} for a particular boundary condition and zero temperature.

Although the authors of \cite{Fuk} proposed four ways for the theoretical derivation of the chiral magnetic effect in unbounded space, there remained some uncertainties as to whether it emerges in thermal equilibrium, or is exclusively a non-equilibrium phenomenon, see \cite{Khar,Yam,Ban}. A recent assertion in \cite{Ben} states that, in a rigorous field-theoretical treatment, the chiral magnetic effect, unlike the chiral separation one, is absent in thermal equilibrium. In my opinion, this issue needs to be clarified, and that is why I consider a general quantum system with Dirac fermions and with the only restriction that is due to chiral symmetry:
\begin{equation}\label{1.1}
\left[H,\gamma^5\right]_{-}=0;
\end{equation}
here $H$ is the one-paticle Dirac Hamiltonian operator and $\gamma^5=-{\rm i}\gamma^0\gamma^1\gamma^2\gamma^3$. I shall develop a field-theoretical approach basing on the path integral formalism for Dirac fermions at non-zero temperature, as well as at finite conventional and chiral chemical potentials, and find the persisting chiral currents that are generated in such a system in thermal equilibrium. The influence of the explicit violation of chiral symmetry by the fermion mass term on the currents will be studied. The developed approach is directly applicable to quantum fermionic systems in the background of a static magnetic field. There are two distinct configurations of magnetic field, covering all plausible ones: the uniform one with the infinite total flux and the nonuniform one with a finite total flux. I shall find out that both the chiral magnetic and chiral separation effects in thermal equilibrium are substantiated on equally the same footing, being independent of the chiral symmetry violating mass.

In the next section elements of the path integral formalism for quantum systems with chiral symmetry are introduced, and generating functionals for the chiral currents are derived and analysed. In section 3 the persisting chiral currents are considered in general and in the uniform magnetic field background. The influence of the explicit violation of chiral symmetry by fermion mass is studied in section 4. The persisting chiral currents in the nonuniform magnetic field background with finite flux are considered in section 5. The results are discussed and summarized in section 6.

\section{Path integral formalism for quantum systems with chiral symmetry}
\setcounter{equation}{0}

Let us consider the following integral over Grassmann fields in Euclidean space-time ($\boldsymbol{x},\tau$):
\begin{equation}\label{2.1}
{\rm e}^{-\beta \Gamma_{\mu_{5}}[\boldsymbol{\eta}]}=C \int  D\psi^{\dagger}D\psi \, 
{\rm exp}\left\{-\int\limits_{0}^{\beta}d\tau \int d^{3}x \, \psi^{\dagger}\left[\partial_{\tau}+H-\mu_{5}\gamma^{5}+\gamma^{0}\boldsymbol{\gamma}\cdot\boldsymbol{\eta}\left(\boldsymbol{x}\right)\right]\psi\right\}, 
\end{equation}
with $H$ satisfying \eqref{1.1} and the Grassmann fields obeying the antiperiodicity condition at the endpoints of the imaginary time interval,
\begin{equation}\label{2.2}
\left.\psi(\boldsymbol{x},\tau)\right|_{\tau=\beta}=-\left.\psi(\boldsymbol{x},\tau)\right|_{\tau=0}, \, \, \, 
\left.\psi^{\dagger}(\boldsymbol{x},\tau)\right|_{\tau=\beta}=-\left.\psi^{\dagger}(\boldsymbol{x},\tau)\right|_{\tau=0};
\end{equation}
here natural units ${\hbar} = c = k_{B} = 1$ are used, $C$ is a (divergent in the continuum limit) normalization constant, 
$\mu_5$ is the chiral (or, to be more precise, axial) chemical potential, $\beta = T^{-1}$, and $T$ is the equilibrium temperature. Functional 
$\Gamma_{\mu_{5}}[\boldsymbol{\eta}]$ is the generating functional for a current which is induced in thermal equilibrium:
\begin{equation}\label{2.3}
\left\langle \boldsymbol{J}(\boldsymbol{y})\right\rangle_{\beta,\mu_{5}}=\left.\frac{\delta \Gamma_{\mu_{5}}[\boldsymbol{\eta}]}{\delta \boldsymbol{\eta}(\boldsymbol{y})}\right|_{\boldsymbol{\eta}=0} = \frac{\int D\psi^{\dagger}D\psi \, {\rm e}^{-S(\mu_5)}\, \int\limits_{0}^{\beta}d\tau' \psi^{\dagger}(\boldsymbol{y},\tau')\gamma^{0}\boldsymbol{\gamma}\psi(\boldsymbol{y},\tau')}{{\beta} \int D\psi^{\dagger}D\psi \, {\rm e}^{-S(\mu_5)}},
\end{equation}
where
\begin{equation}\label{2.4}
S(\mu_{5})=\int\limits_{0}^{\beta}d\tau \int d^{3}x \, \psi^{\dagger}(\partial_{\tau}+H-\mu_5\gamma^{5})\psi.
\end{equation}
Integration over the Grassmann fields in \eqref{2.1} yields in a standard way
\begin{eqnarray}\label{2.5}
\Gamma_{\mu_{5}}[\boldsymbol{\eta}]=-\frac{1}{\beta}{\rm ln}{\rm Det}\left(\partial_{\tau}+H-\mu_{5}\gamma^{5}+\gamma^{0}\boldsymbol{\gamma}\cdot \boldsymbol{\eta}\right) \nonumber \\ 
=-\frac{1}{\beta}\int\limits_{0}^{\beta}d\tau\int d^{3}x \, {\rm tr}\left\langle \left.\boldsymbol{x},\tau\right|\ln\left.\left(\partial_{\tau}+H-\mu_{5}\gamma^{5}+\gamma^{0}\boldsymbol{\gamma}\cdot\boldsymbol{\eta}\right)\right|\tau,\boldsymbol{x}\right\rangle,
\end{eqnarray}
where the trace over spinor indices is denoted by ${\rm tr}$. If both $H$ and $\mu_{5}$ are independent of $\tau$, then integration over $\tau$ is explicitly performed, yielding
\begin{equation}\label{2.6}
\Gamma_{\mu_{5}}[\boldsymbol{\eta}]=-\frac{1}{\beta}\sum\limits_{n\in \mathbb{Z}}\int d^{3}x \, {\rm tr}\left\langle \left.\boldsymbol{x}\right|\ln\left.\left(H-\mu_{5}\gamma^{5}+\gamma^{0}\boldsymbol{\gamma}\cdot\boldsymbol{\eta}-{\rm i}\omega_n\right)\right|\boldsymbol{x}\right\rangle,
\end{equation}
where $\omega_n=(2n+1)\pi/\beta$ as a consequence of condition \eqref{2.2} and 
$\mathbb{Z}$ is the set of integer numbers. Defining zeta function
\begin{equation}\label{2.7}
\zeta_{s, \mu_{5}}\left[\boldsymbol{\eta}\right]=\frac{\lambda^{s}}{\beta}
\sum\limits_{n\in \mathbb{Z}}\int d^{3}x \, {\rm tr}\left\langle  \boldsymbol{x}\right|\left(H-\mu_{5}\gamma^{5}+\gamma^{0}\boldsymbol{\gamma}\cdot\boldsymbol{\eta}-{\rm i}\omega_n\right)^{-s}\left|\boldsymbol{x}\right\rangle,
\end{equation}
where $\lambda$ is a parameter of dimension of momentum, one gets relation
$$
\Gamma_{\mu_{5}}[\boldsymbol{\eta}]=\frac{d}{ds}\left.\zeta_{s, \mu_{5}}\left[\boldsymbol{\eta}\right]\right|_{s=0}.
$$
But instead of dealing with \eqref{2.7}, it is more efficient to perform summation over $n$ directly in \eqref{2.6}, see Appenix B in \cite{SiV8}. As a result we get 
\begin{equation}\label{2.8}
\Gamma_{\mu_{5}}[\boldsymbol{\eta}]=-\frac{1}{\beta}
\int d^{3}x \, {\rm tr}\left\langle\boldsymbol{x}\right|
\ln\cosh\left[\frac{\beta}{2}\left(H-\mu_{5}\gamma^{5}+\gamma^{0}
\boldsymbol{\gamma}\cdot\boldsymbol{\eta}\right)\right]\left|
\boldsymbol{x}\right\rangle.
\end{equation}

In view of chiral symmetry \eqref{1.1}, let us choose a common system of eigenfunctions for $H$ and $\gamma^5$: 
\begin{equation}\label{2.9}
\gamma^{5}\left\langle \pm, \boldsymbol{x} \right|=\pm\left\langle \pm, \boldsymbol{x} \right|, \, \,  H\left\langle \pm, \boldsymbol{x}\right|=E\left\langle \pm, \boldsymbol{x}\right| .
\end{equation}
Thus, using the chiral representation for Dirac matrices, 
\begin{equation}\label{2.10}
\gamma^0=\begin{pmatrix}
0\,&\,I\\
I\,&\,0
\end{pmatrix},\qquad
\boldsymbol{\gamma}=\begin{pmatrix}
0\,&-\boldsymbol{\sigma}\\
\boldsymbol{\sigma}&\,0
\end{pmatrix}, \quad 
\gamma^5=\begin{pmatrix}
-I\,&\,0\\
0\,&\,I
\end{pmatrix}
\end{equation}
($\sigma^1$, $\sigma^2$, and $\sigma^3$ are the Pauli matrices), we obtain
\begin{equation}\label{2.11}
\Gamma_{\mu_{5}}[\boldsymbol{\eta}]=-\frac{1}{\beta}\sum\limits_{\pm}\int d^{3}x \, {\rm tr}\left\langle \pm,\boldsymbol{x}\right|\ln\cosh\left[\frac{\beta}{2}\left(H_{\pm}\mp \mu_5\mp\boldsymbol{\sigma}\cdot \boldsymbol{\eta} \right)\right]\left|\boldsymbol{x},\pm \right\rangle,
\end{equation}
where Hamiltonian $H$ is splitted as
\begin{equation}\label{2.12}
H=\begin{pmatrix}
H_{-}\,&\,0\\
0\,&\,H_{+}
\end{pmatrix}.
\end{equation}

Right now we can make some qualitative conclusions. As a consequence of \eqref{2.9} and \eqref{2.12}, spectra of $H_+$ and $H_-$ coincide. Therefore in \eqref{2.11} we are dealing with expression
$$
\sum\limits_{\pm}\ln\cosh\left[\frac{\beta}{2}\left(E\mp\mu_5\mp\boldsymbol{\sigma}\cdot \boldsymbol{\eta}\right)\right],
$$
which reduces to 
$$
\ln\left\{\cosh(\beta E)+\cosh\left[\beta\left(\mu_{5}+\boldsymbol{\sigma}\cdot \boldsymbol{\eta}\right)\right]\right\}-\ln 2,
$$
and the latter clearly contains linear in $\mu_5\boldsymbol{\sigma}\cdot \boldsymbol{\eta}$ terms. As to the functional with conventional chemical potential $\mu$, $\Gamma_{\mu}[\boldsymbol{\eta}]$, which is defined by relation
\begin{equation}\label{2.13}
{\rm e}^{-\beta \Gamma_{\mu}[\boldsymbol{\eta}]}=C\int D\psi^{\dagger}D\psi \, 
{\rm exp}\left\{-\int\limits_{0}^{\beta}d\tau \int d^{3}x \, \psi^{\dagger}\left[\partial_{\tau}+H-\mu+\gamma^{0}\boldsymbol{\gamma}\cdot \boldsymbol{\eta}\left(\boldsymbol{x}\right)\right]\psi\right\},
\end{equation}
it fails to produce linear in $\boldsymbol{\eta}$ terms. Really, functional $\Gamma_{\mu}[\boldsymbol{\eta}]$ involves expression
$$
\ln\left\{\cosh\left[\beta(E-\mu)\right]+\cosh(\beta\boldsymbol{\sigma}\cdot\boldsymbol{\eta})\right\}-\ln 2,
$$
which clearly contains even powers of $|\boldsymbol{\eta}|$ only. Let us further consider functionals $\Gamma_{\mu_{5}}[{\boldsymbol{\eta}}_{5}]$ and $\Gamma_{\mu}[{\boldsymbol{\eta}}_{5}]$ which are defined by relations
\begin{equation}\label{2.14}
{\rm e}^{-\beta \Gamma_{\mu_{5}}[{\boldsymbol{\eta}}_{5}]}=C\int D\psi^{\dagger}D\psi \, 
{\rm exp}\left\{-\int\limits_{0}^{\beta}d\tau \int d^{3}x \, \psi^{\dagger}\left[\partial_{\tau}+H-\mu_{5}{\gamma}^{5}+\gamma^{0}\boldsymbol{\gamma}{\gamma}^{5}\cdot\boldsymbol{\eta}_{5}\left(\boldsymbol{x}\right)\right]\psi\right\},
\end{equation}
and
\begin{equation}\label{2.15}
{\rm e}^{-\beta \Gamma_{\mu}[{\boldsymbol{\eta}}_{5}]}=C\int D\psi^{\dagger}D\psi \, 
{\rm exp}\left\{-\int\limits_{0}^{\beta}d\tau \int d^{3}x \, \psi^{\dagger}\left[\partial_{\tau}+H-\mu+\gamma^{0}\boldsymbol{\gamma}\gamma^{5}\cdot \boldsymbol{\eta}_5\left(\boldsymbol{x}\right)\right]\psi\right\}.
\end{equation}
In a similar to the above way, one can show that we are dealing with expression
$$
\sum\limits_{\pm}\ln \cosh\left[\frac{\beta}{2}\left(E \mp \mu_{5}-\boldsymbol{\sigma}\cdot\boldsymbol{\eta}_{5}\right)\right]=\ln\left\{\cosh\left[\beta\left(E-\boldsymbol{\sigma}\cdot\boldsymbol{\eta}_{5}\right)\right]+\cosh(\beta{\mu}_{5})\right\}-\ln 2
$$
in the case of $\Gamma_{\mu_{5}}[{\boldsymbol{\eta}}_{5}]$, and with expression
$$
2\ln \cosh\left[\frac{\beta}{2}\left(E-\mu-\boldsymbol{\sigma}\cdot \boldsymbol{\eta}_{5}\right)\right]
$$
in the case of $\Gamma_{\mu}[{\boldsymbol{\eta}}_{5}]$. Clearly functional $\Gamma_{\mu_{5}}[{\boldsymbol{\eta}}_{5}]$ involves linear in $E\boldsymbol{\sigma}\cdot {\boldsymbol{\eta}}_{5}$ terms, which are canceled upon summation over the energy sign. On the contrary, functional $\Gamma_{\mu}[{\boldsymbol{\eta}}_{5}]$ involves, in addition, linear in 
$\mu\boldsymbol{\sigma}\cdot \boldsymbol{\eta}_5$ terms. We thus see that, among the four above functionals, only $\Gamma_{\mu_{5}}[\boldsymbol{\eta}]$ and $\Gamma_{\mu}[{\boldsymbol{\eta}}_{5}]$ are linearly dependent on the variation, 
$\boldsymbol{\eta}$ and 
${\boldsymbol{\eta}}_{5}$, correspondingly. This is certainly a necessary, and not sufficient, condition for inducing the appropriate currents, $\left\langle \boldsymbol{J}(\boldsymbol{y})\right\rangle_{\beta,\mu_{5}}$ and $\left\langle \boldsymbol{J}^{5}(\boldsymbol{y})\right\rangle_{\beta,\mu}$. The quantitative results are obtained in the next sections.

\section{Persisting currents in thermal equilibrium}
\setcounter{equation}{0}

Using representation \eqref{2.11}, we obtain the following expression for current \eqref{2.3} (changing 
$\boldsymbol{y}$ to $\boldsymbol{x}$): 
\begin{equation}\label{3.1}
\left\langle \boldsymbol{J}(\boldsymbol{x})\right\rangle_{\beta,\mu_5}=\frac{1}{2}\sum\limits_{\pm}(\pm){\rm tr}\boldsymbol{\sigma}\left\langle \pm, \boldsymbol{x}\right|\tanh \left[\frac{\beta}{2}\left(H_{\pm}\mp \mu_5\right)\right]\left|\boldsymbol{x}, \pm \right\rangle.
\end{equation}

If $H$ is an unbounded self-adjoint operator, then its resolvent, $(H-\omega)^{-1}$, is defined at ${\rm Im} \, \omega \neq 0$, see, e.g., \cite{Reed}. The following relation is valid
\begin{equation}\label{3.2}
\Upsilon \, f(H) = \int\limits_{\Omega}\frac{d\omega}{2\pi{\rm i}} \Upsilon
 \, (H-\omega)^{-1} f(\omega),
\end{equation}
where $\Upsilon$ is a matrix (element of the Clifford algebra), $f(H)$ is a function of operator $H$, and $\Omega$ is the contour consisting of two straight antiparallel lines, $(-\infty +{\rm i}\epsilon,\ +\infty +{\rm i}\epsilon)$ and $(+\infty -{\rm i}\epsilon,\ -\infty -{\rm i}\epsilon)$, in the complex $\omega$ plane. Taking the trace of matrix elements of \eqref{3.2} and diminishing a distance between the lines, $2\epsilon$, to zero, one obtains relation
\begin{equation}\label{3.3}
{\rm tr}\Upsilon\left\langle \left.\boldsymbol{x}\right|\right.f(H)\left|\left.\boldsymbol{x}\right.\right\rangle=\int\limits_{-\infty}^{\infty}dE \, \tau_{\Upsilon}(\boldsymbol{x}, E)f(E),
\end{equation}
where
\begin{multline}\label{3.4}
\tau_{\Upsilon}(\boldsymbol{x}, E)=\frac{1}{2\pi{\rm i}}\lim\limits_{\epsilon\rightarrow 0} \, 
{\rm tr}\Upsilon\left\langle \left.\boldsymbol{x}\right|\right.[(H-E-{\rm i}\epsilon)^{-1}-(H-E+{\rm i}\epsilon)^{-1}]
\left|\left.\boldsymbol{x}\right.\right\rangle \\
={\rm tr}\Upsilon\left\langle \left.\boldsymbol{x}\right|\right.\delta(H-E)\left|\left.\boldsymbol{x}\right.\right\rangle
\end{multline}
is the local spectral density weighted with $\Upsilon$. Hence, current \eqref{3.1} can be presented as
\begin{equation}\label{3.5}
\left\langle \boldsymbol{J}(\boldsymbol{x})\right\rangle_{\beta,\mu_5}=\frac{1}{2}\int\limits_{-\infty}^{\infty}dE\sum\limits_{\pm}(\pm)
\boldsymbol{\tau}_{\pm}(\boldsymbol{x}, E)\tanh\left[\frac{\beta}{2}\left(E\mp\mu_5\right)\right],
\end{equation}
where
\begin{equation}\label{3.6}
\boldsymbol{\tau}_{\pm}(\boldsymbol{x}, E)=
{\rm tr}\boldsymbol{\sigma}\left\langle \left.\pm,\boldsymbol{x}\right|\right.\delta(H_{\pm}-E)\left|\left.\boldsymbol{x},\pm\right.\right\rangle.
\end{equation}
Note that the axial current which is obtained by variation of the functional defined by \eqref{2.15} can be presented in a similar way:
\begin{equation}\label{3.7}
\left\langle \boldsymbol{J}^5(\boldsymbol{x})\right\rangle_{\beta,\mu}=\frac{1}{2}\int\limits_{-\infty}^{\infty}dE\sum\limits_{\pm}
\boldsymbol{\tau}_{\pm}(\boldsymbol{x}, E)\tanh\left[\frac{\beta}{2}\left(E-\mu\right)\right].
\end{equation}

Considering a quantum massless fermionic system in the background of magnetic field $\boldsymbol{B}=\boldsymbol{\partial}\times\boldsymbol{A}(\boldsymbol{x})$ that is static and uniform, the Dirac Hamiltonian operator takes form
\begin{equation}\label{3.8}
H=-{\rm i}\gamma^{0}\boldsymbol{\gamma}\cdot\left[\boldsymbol{\partial}-{\rm i}e\boldsymbol{A}(\boldsymbol{x})\right].
\end{equation}
Solutions to the appropriate Dirac equation are well described in the literature, see, e.g., \cite{Akhie}. The energy spectrum is given by
\begin{eqnarray}\label{3.9}
E_{lk}=\left\{\begin{array}{l} \sqrt{2l|eB|+k^{2}}\\ [6 mm]
- \sqrt{2l|eB|+k^{2}}\end{array} \right\},
\quad -\infty<k<\infty, \quad l=0,1,2,...\, ,
\end{eqnarray}
$k$ is the value of the wave number vector along the magnetic field, and $l$ enumerates the Landau levels. 
Unlike the lowest ($l=0$) one, the levels with $l\geq 1$ are doubly degenerate: 
$\left\langle \pm, \boldsymbol{x}, {\rm up} \, |l,q,k\right\rangle$ and $\left\langle \pm, \boldsymbol{x}, {\rm down} \, |l, q, k\right\rangle$; they correspond to opposite spin projections on the direction of the magnetic field. Choosing gauge $\boldsymbol{A}=(-x^{2}B, 0, 0)$ with the magnetic field along the $x^{3}$ axis, $\boldsymbol{B}=(0, 0, B)$, one obtains the following expressions for the modes in the case of $eB>0$:
\begin{eqnarray}
\left.\left\langle \pm, \boldsymbol{x}, {\rm up} \, |l,q,k\right\rangle\right|_{E_{lk}>0}=\frac{C_{\pm}^{(l)}}{4\pi}{\rm e}^{{\rm i}(qx^1+kx^3)}\left(\begin{array}{l} \left(1\mp\frac{k}{E_{lk}}\right)Y_l^q(x^2)\\ [6 mm]
\mp\frac{\sqrt{2leB}}{E_{lk}}Y_{l-1}^q(x^2)\end{array} \right),
\label{3.10}
\end{eqnarray}
\begin{eqnarray}
\left.\left\langle \pm, \boldsymbol{x}, {\rm up} \, |l,q,k\right\rangle\right|_{E_{lk}<0}=\frac{C_{\pm}^{(l)}}{4\pi}{\rm e}^{-{\rm i}(qx^1+kx^3)}\left(\begin{array}{l} \mp\left(1\pm\frac{k}{E_{lk}}\right)Y_l^{-q}(x^2)\\ [6 mm]
\frac{\sqrt{2leB}}{E_{lk}}Y_{l-1}^{-q}(x^2)\end{array} \right),
\label{3.11}
\end{eqnarray}
\begin{eqnarray}
\left.\left\langle \pm, \boldsymbol{x}, {\rm down} \, |l,q,k\right\rangle\right|_{E_{lk}>0}=\frac{{\tilde C}_{\pm}^{(l)}}{4\pi}{\rm e}^{{\rm i}(qx^1+kx^3)}\left(\begin{array}{l} \mp\frac{\sqrt{2leB}}{E_{lk}}Y_l^q(x^2)\\ [6 mm]
\left(1\pm\frac{k}{E_{lk}}\right)Y_{l-1}^q(x^2)\end{array} \right),
\label{3.12}
\end{eqnarray}
\begin{eqnarray}
\left.\left\langle \pm, \boldsymbol{x}, {\rm down} \, |l,q,k\right\rangle\right|_{E_{lk}<0}=\frac{{\tilde C}_{\pm}^{(l)}}{4\pi}{\rm e}^{-{\rm i}(qx^1+kx^3)}\left(\begin{array}{l} \frac{\sqrt{2leB}}{E_{lk}}Y_l^{-q}(x^2)\\ [6 mm]
\mp\left(1\mp\frac{k}{E_{lk}}\right)Y_{l-1}^{-q}(x^2)\end{array} \right),
\label{3.13}
\end{eqnarray}
\begin{eqnarray}
\left.\left\langle \pm, \boldsymbol{x} \, |0,q,k\right\rangle\right|_{E_{0k}>0}=\frac{C_{\pm}^{(0)}}{2\pi}{\rm e}^{{\rm i}(qx^1+kx^3)}\left(\begin{array}{l} \Theta(\mp k)Y_0^q(x^2)\\ [6 mm]
 \quad \quad 0 \end{array} \right),
\label{3.14}
\end{eqnarray}
and
\begin{eqnarray}
\left.\left\langle \pm, \boldsymbol{x} \, |0,q,k\right\rangle\right|_{E_{0k}<0}=\frac{C_{\pm}^{(0)}}{2\pi}{\rm e}^{-{\rm i}(qx^1+kx^3)}\left(\begin{array}{l} \mp\Theta(\mp k)Y_0^{-q}(x^2)\\ [6 mm]
 \quad \quad 0 \end{array} \right);
\label{3.15}
\end{eqnarray}
here $-\infty<q<\infty$, $|C_{\pm}^{(l)}|=|{\tilde C}_{\pm}^{(l)}|=1$, $\Theta(u)$ is the unit step function, and
$$
Y_l^q(u)=\sqrt{\frac{(eB)^{1/2}}{2^ll!\pi^{1/2}}}\exp\left[-\frac{eB}{2}\left(u+\frac{q}{eB}\right)^2\right]H_l\left(\sqrt{eB}u+\frac{q}{\sqrt{eB}}\right),
$$
$H_l(v)=(-1)^l {\rm e}^{v^2}\frac{{\rm d}^l}{{\rm d}v^l}{\rm e}^{-v^2}$ is the Hermite polynomial. Function $Y_l^q(u)$ obeys 
condition 
\begin{equation}\label{3.16}
\int\limits_{-\infty}^{\infty}{\rm d}q\,Y_l^{q}(u)Y_{l'}^{q}(u)=eB \,\delta_{ll'}.
\end{equation}
The modes in the case of $eB<0$ are obtained by charge conjugation, i.e., going to $eB \rightarrow -eB$ and multiplying the complex conjugates of the above modes by ${\rm i}\gamma^{2}$ (the energy sign is reversed). 

In view of the the nondiagonal form of $\sigma^1$ and 
$\sigma^2$ and orthogonality of $Y_{l-1}^q$ to $Y_l^q$, we obtain
\begin{equation}\label{3.17}
\tau^{1}_{\pm}(\boldsymbol{x}, E)=\tau_{\pm}^{2}(\boldsymbol{x}, E)=0.
\end{equation}
As to $\tau^{3}_{\pm}(\boldsymbol{x}, E)$, the contribution of ``up'' and ``down'' modes is canceled owing to relation
\begin{multline}\label{3.18}
\int\limits_{-\infty}^{\infty}{\rm d}q \, \left[\left\langle k,q,l\, | \, {\rm up}, \boldsymbol{x}, \pm \right\rangle \sigma^3 \left\langle \pm, \boldsymbol{x}, {\rm up} \, | \, l,q,k\right\rangle \right. \\
\left.+ \left\langle k,q,l \, | \, {\rm down}, \boldsymbol{x}, \pm \right\rangle \sigma^3\left\langle \pm, \boldsymbol{x}, {\rm down} \, | \, l,q,k\right\rangle \right] \sim k,
\end{multline}
that vanishes upon summation over the sign of $k$, and only the lowest Landau level contributes, yielding
\begin{equation}\label{3.19}
\tau^{3}_{\pm}(\boldsymbol{x}, E)=\frac{eB}{(2\pi)^{2}}.
\end{equation}
The same is obtained in the case of $eB<0$ also. 

Inserting \eqref{3.19} into \eqref{3.5} and \eqref{3.7}, we obtain
\begin{equation}\label{3.20}
\left\langle J^{3}\right\rangle_{\beta, \mu_{5}}=-\frac{eB}{2\pi^{2}}\mu_{5},
\end{equation}
that is the chiral magnetic effect \cite{Fuk}, and
\begin{equation}\label{3.21}
\left\langle J^{35}\right\rangle_{\beta, \mu}=-\frac{eB}{2\pi^{2}}\mu,
\end{equation}
that is the chiral separation effect \cite{Met}. Thus, both effects, (3.20) and (3.21), are substantiated on the same grounds.

The results can be extended to a more general setup with both chemical potentials, $\mu$ and $\mu_5$, being nonvanishing. We define
\begin{multline}\label{3.22}
\Gamma[\boldsymbol{\eta}, \boldsymbol{\eta}_{5}]=-\frac{1}{\beta}\ln\Biggl(\int D\psi^{\dagger}D\psi \Biggr. \\ 
\Biggl. \times {\rm exp}\left\{-\int\limits_{0}^{\beta}d\tau \int d^{3}x \, \psi^{\dagger}\left[\partial_{\tau}+H-\mu-\mu_{5}\gamma^{5}+\gamma^{0}\boldsymbol{\gamma}\cdot\boldsymbol{\eta}\left(\boldsymbol{x}\right)+\gamma^{0}\boldsymbol{\gamma}\gamma^{5}\cdot\boldsymbol{\eta}_5\left(\boldsymbol{x}\right)\right]\psi\right\}\Biggr), 
\end{multline}
as well as
\begin{equation}\label{3.23}
\left\langle \boldsymbol{J}(\boldsymbol{y})\right\rangle_{\beta}=\left.\frac{\delta \Gamma[\boldsymbol{\eta}, \boldsymbol{\eta}_5]}{\delta \boldsymbol{\eta}(\boldsymbol{y})}\right|_{\boldsymbol{\eta}=\boldsymbol{\eta}_5=0} 
\end{equation}
and
\begin{equation}\label{3.24}
\left\langle \boldsymbol{J}^5(\boldsymbol{y})\right\rangle_{\beta}=\left.\frac{\delta \Gamma[\boldsymbol{\eta}, \boldsymbol{\eta}_5]}{\delta \boldsymbol{\eta}_5(\boldsymbol{y})}\right|_{\boldsymbol{\eta}=\boldsymbol{\eta}_5=0}. 
\end{equation}
Similar to the above, we obtain
\begin{equation}\label{3.25}
\left\langle \boldsymbol{J}(\boldsymbol{x})\right\rangle_{\beta}=\frac{1}{2}\int\limits_{-\infty}^{\infty}dE\sum\limits_{\pm}(\pm)
\boldsymbol{\tau}_{\pm}(\boldsymbol{x}, E)\tanh\left[\frac{\beta}{2}\left(E-\mu\mp\mu_5\right)\right]
\end{equation}
and
\begin{equation}\label{3.26}
\left\langle \boldsymbol{J}^5(\boldsymbol{x})\right\rangle_{\beta}=\frac{1}{2}\int\limits_{-\infty}^{\infty}dE\sum\limits_{\pm}
\boldsymbol{\tau}_{\pm}(\boldsymbol{x}, E)\tanh\left[\frac{\beta}{2}\left(E-\mu\mp\mu_5\right)\right].
\end{equation}

In the end of the previous section, while discussing the dependence of functional 
$\Gamma_{\mu}[\boldsymbol{\eta}]$ on $\boldsymbol{\eta}$, we implicitly assumed that 
\begin{equation}\label{3.27}
\boldsymbol{\tau}_{\pm}(\boldsymbol{x}, E)=\boldsymbol{\tau}(\boldsymbol{x}, E).
\end{equation}
In view of this, we get
\begin{equation}\label{3.28}
\left\langle \boldsymbol{J}(\boldsymbol{x})\right\rangle_{\beta}= -\sinh\left(\beta\mu_5\right)\int\limits_{-\infty}^{\infty}dE \frac{\boldsymbol{\tau}(\boldsymbol{x}, E)}{\cosh\left[\beta\left(E-\mu\right)\right]+\cosh\left(\beta\mu_5\right)} 
\end{equation}
and
\begin{equation}\label{3.29}
\left\langle \boldsymbol{J}^5(\boldsymbol{x})\right\rangle_{\beta}=\int\limits_{-\infty}^{\infty}dE \frac{\boldsymbol{\tau}(\boldsymbol{x}, E)\sinh\left[\beta\left(E-\mu\right)\right]}{\cosh\left[\beta\left(E-\mu\right)\right]+\cosh\left(\beta\mu_5\right)}.
\end{equation}
As is clear from \eqref{3.28}, current $\left\langle \boldsymbol{J}(\boldsymbol{x})\right\rangle_{\beta}$ vanishes at $\mu_5=0$. While discussing the dependence of functional 
$\Gamma_{\mu_{5}}[{\boldsymbol{\eta}}_{5}]$ on ${\boldsymbol{\eta}}_{5}$, we implicitly assumed that 
\begin{equation}\label{3.30}
\boldsymbol{\tau}_{+}(\boldsymbol{x}, E)=\boldsymbol{\tau}_{-}(\boldsymbol{x}, -E).
\end{equation} 
Hence, we get in the latter case
\begin{multline}\label{3.31}
\left\langle \boldsymbol{J}(\boldsymbol{x})\right\rangle_{\beta}= \int\limits_{0}^{\infty}dE \left\{\frac{\boldsymbol{\tau}_{+}(\boldsymbol{x}, E)\sinh\left[\beta\left(E-\mu_5\right)\right]}{\cosh\left[\beta\left(E-\mu_5\right)\right]+\cosh\left(\beta\mu\right)} \right. \\
\left. - \frac{\boldsymbol{\tau}_{-}(\boldsymbol{x}, E)\sinh\left[\beta\left(E+\mu_5\right)\right]}{\cosh\left[\beta\left(E+\mu_5\right)\right]+\cosh\left(\beta\mu\right)}
\right\}
\end{multline}
and
\begin{multline}\label{3.32}
\left\langle \boldsymbol{J}^5(\boldsymbol{x})\right\rangle_{\beta}= -\sinh\left(\beta\mu\right)\int\limits_{0}^{\infty}dE \left\{\frac{\boldsymbol{\tau}_{+}(\boldsymbol{x}, E)}{\cosh\left[\beta\left(E-\mu_5\right)\right]+\cosh\left(\beta\mu\right)} \right. \\
\left. + \frac{\boldsymbol{\tau}_{-}(\boldsymbol{x}, E)}{\cosh\left[\beta\left(E+\mu_5\right)\right]+\cosh\left(\beta\mu\right)}
\right\}.
\end{multline}
Clearly, current $\left\langle \boldsymbol{J}^5(\boldsymbol{x})\right\rangle_{\beta}$  vanishes at $\mu=0$.

In the case when both conditions \eqref{3.27} and \eqref{3.30} are satisfied, i.e., 
$\boldsymbol{\tau}(\boldsymbol{x}, E)$ is an even function of $E$, one obtains
\begin{multline}\label{3.33}
\left\langle \boldsymbol{J}(\boldsymbol{x})\right\rangle_{\beta}= -2\sinh\left(\beta\mu_5\right)\int\limits_{0}^{\infty}dE \, \boldsymbol{\tau}(\boldsymbol{x}, E) \\
\times \frac{\cosh\left(\beta E\right)\cosh\left(\beta\mu\right)+\cosh\left(\beta\mu_5\right)}{\sinh^2\left(\beta E\right)+2\cosh\left(\beta E\right)\cosh\left(\beta\mu\right)\cosh\left(\beta\mu_5\right)+
	\cosh^2\left(\beta\mu\right)+\cosh^2\left(\beta\mu_5\right)} 
\end{multline}
and
\begin{multline}\label{3.34}
\left\langle \boldsymbol{J}^5(\boldsymbol{x})\right\rangle_{\beta}= -2\sinh\left(\beta\mu\right)\int\limits_{0}^{\infty}dE \, \boldsymbol{\tau}(\boldsymbol{x}, E) \\
\times \frac{\cosh\left(\beta E\right)\cosh\left(\beta\mu_5\right)+\cosh\left(\beta\mu\right)}{\sinh^2\left(\beta E\right)+2\cosh\left(\beta E\right)\cosh\left(\beta\mu\right)\cosh\left(\beta\mu_5\right)+
	\cosh^2\left(\beta\mu\right)+\cosh^2\left(\beta\mu_5\right)}. 
\end{multline}
Current \eqref{3.33} turns into current \eqref{3.34} under interchange of $\mu$ and $\mu_5$. Defining the left and right chiral currents, 
\begin{equation}\label{3.35}
\left\langle \boldsymbol{J}^{L}(\boldsymbol{x})\right\rangle_{\beta} = \frac12 \left(\left\langle \boldsymbol{J}(\boldsymbol{x})\right\rangle_{\beta}+\left\langle \boldsymbol{J}^{5}(\boldsymbol{x})\right\rangle_{\beta}\right)  
\end{equation}
and
\begin{equation}\label{3.36}
\left\langle \boldsymbol{J}^{R}(\boldsymbol{x})\right\rangle_{\beta} = \frac12 \left(\left\langle \boldsymbol{J}(\boldsymbol{x})\right\rangle_{\beta}-\left\langle \boldsymbol{J}^{5}(\boldsymbol{x})\right\rangle_{\beta}\right),  
\end{equation}
one obtains
\begin{multline}\label{3.37}
\left\langle \boldsymbol{J}^L(\boldsymbol{x})\right\rangle_{\beta}= -\frac12\int\limits_{0}^{\infty}dE \, \boldsymbol{\tau}(\boldsymbol{x}, E)  \\
\times \frac{2\cosh\left(\beta E\right)\sinh\left[\beta\left(\mu+\mu_5\right)\right] + \sinh\left(2\beta\mu_5\right)}{\sinh^2\left(\beta E\right)+2\cosh\left(\beta E\right)\cosh\left(\beta\mu\right)\cosh\left(\beta\mu_5\right)+
	\cosh^2\left(\beta\mu\right)+\cosh^2\left(\beta\mu_5\right)} 
\end{multline}
and
\begin{multline}\label{3.38}
\left\langle \boldsymbol{J}^R(\boldsymbol{x})\right\rangle_{\beta}= \frac12\int\limits_{0}^{\infty}dE \, \boldsymbol{\tau}(\boldsymbol{x}, E)  \\
\times \frac{2\cosh\left(\beta E\right)\sinh\left[\beta\left(\mu-\mu_5\right)\right] - \sinh\left(2\beta\mu\right)}{\sinh^2\left(\beta E\right)+2\cosh\left(\beta E\right)\cosh\left(\beta\mu\right)\cosh\left(\beta\mu_5\right)+
	\cosh^2\left(\beta\mu\right)+\cosh^2\left(\beta\mu_5\right)}. 
\end{multline}

In the case of a static uniform magnetic field strength along the $x^3$ axis, both conditions \eqref{3.27} and \eqref{3.30} are satisfied; moreover, $\boldsymbol{\tau}(\boldsymbol{x}, E)$ is independent of both $\boldsymbol{x}$ and $E$. Inserting \eqref{3.17} and \eqref{3.19} into \eqref{3.33} and \eqref{3.34}, one gets
\begin{equation}\label{3.39}
\left\langle J^{3}\right\rangle_{\beta}=-\frac{eB}{2\pi^{2}}\mu_{5}
\end{equation}
and
\begin{equation}\label{3.40}
\left\langle J^{35}\right\rangle_{\beta}=-\frac{eB}{2\pi^{2}}\mu,
\end{equation}
that coincides with \eqref{3.20} and \eqref{3.21}. The left and right chiral currents take form 
\begin{equation}\label{3.41}
\left\langle J^{3L}\right\rangle_{\beta} =-\frac{e B}{2\pi^2}{\mu}_L  
\end{equation}
and
\begin{equation}\label{3.42}
\left\langle J^{3R}\right\rangle_{\beta} =\frac{e B}{2\pi^2}{\mu}_R,   
\end{equation}
where
\begin{equation}\label{3.43}
{\mu}_L = \frac12(\mu + {\mu}_5) 
\end{equation}
and
\begin{equation}\label{3.44}
{\mu}_R = \frac12(\mu - {\mu}_5) 
\end{equation}
are the left and right chiral chemical potentials. Relation \eqref{3.41} was obtained in \cite{Vil}; perhaps, A.Vilenkin partly anticipated both the chiral magnetic and the chiral separation effects more than 40 years ago.

In a similar to the above way, let us consider the temporal components of currents: 
\begin{equation}\label{3.45}
\left\langle J^0(\boldsymbol{x})\right\rangle_{\beta} = - \frac{1}{2}\int\limits_{-\infty}^{\infty}dE\sum\limits_{\pm}
\tau^0_{\pm}(\boldsymbol{x}, E)\tanh\left[\frac{\beta}{2}\left(E-\mu\mp\mu_5\right)\right]
\end{equation}
and
\begin{equation}\label{3.46}
\left\langle J^{05}(\boldsymbol{x})\right\rangle_{\beta} = - \frac{1}{2}\int\limits_{-\infty}^{\infty}dE\sum\limits_{\pm}
(\pm)\tau^0_{\pm}(\boldsymbol{x}, E)\tanh\left[\frac{\beta}{2}\left(E-\mu\mp\mu_5\right)\right],
\end{equation}
where
\begin{equation}\label{3.47}
\tau^0_{\pm}(\boldsymbol{x}, E)=
{\rm tr} \left\langle \left.\pm,\boldsymbol{x}\right|\right.\delta(H_{\pm}-E)\left|\left.\boldsymbol{x},\pm\right.\right\rangle.
\end{equation}
With the use of \eqref{3.10}-\eqref{3.15} one obtains
\begin{equation}\label{3.48}
\tau^{0}_{\pm}(\boldsymbol{x}, E)=\frac{|eB|}{(2\pi)^{2}}\left[1 + 2\sum\limits_{l=1}^{\infty}\frac{|E| \,  \Theta\left(E^2 - 2l|eB|\right)}{\sqrt{E^2 - 2l|eB|}}\right],
\end{equation}
and, inserting the latter into \eqref{3.45} and \eqref{3.46}, one finally gets 
\begin{multline}\label{3.49}
\left\langle {J}^0\right\rangle_{\beta}=\frac{|eB|}{2\pi^{2}}\Biggl(\mu \Biggr.\\
\Biggl.+ \frac{\beta}{2}\sum\limits_{l=1}^{\infty}\int\limits_{\sqrt{2l|eB|}}^{\infty}dE \sqrt{E^2 - 2l|eB|} \sum\limits_{\pm}\left\{{\rm sech}^2\left[\frac 12\beta\left(E - \mu \mp \mu_5\right)\right] - {\rm sech}^2\left[\frac 12\beta\left(E + \mu \mp \mu_5\right)\right]\right\}\Biggr) 
\end{multline}
and
\begin{multline}\label{3.50}
\left\langle {J}^{05}\right\rangle_{\beta}=\frac{|eB|}{2\pi^{2}}\Biggl(\mu_5 \Biggr.\\
\Biggl.+ \frac{\beta}{2}\sum\limits_{l=1}^{\infty}\int\limits_{\sqrt{2l|eB|}}^{\infty}dE \sqrt{E^2 - 2l|eB|} \sum\limits_{\pm}\left\{{\rm sech}^2\left[\frac 12\beta\left(E \mp \mu -  \mu_5\right)\right] - {\rm sech}^2\left[\frac 12\beta\left(E \mp \mu + \mu_5\right)\right]\right\}\Biggr). 
\end{multline}
Thus, both fermion number density 
$\left\langle {J}^{0}\right\rangle_{\beta}$ and axial charge density 
$\left\langle {J}^{05}\right\rangle_{\beta}$ depend on temperature, and this is due to a contribution of the Landau levels with $l \geq 1$. Note that $\left\langle {J}^{0}\right\rangle_{\beta}$ vanishes at $\mu = 0$, while $\left\langle {J}^{05}\right\rangle_{\beta}$ vanishes at $\mu_5 = 0$. 
Note also that the contribution of the $l \geq 1$ levels decreases exponentially 
in the zero-temperature limit.     

Relations (3.49) and (3.50) take form
\begin{multline}\label{3.51}
\left\langle J^0\right\rangle_{\beta} = \frac{|eB|}{2\pi^{2}} \mu \Biggl\{1 \Biggr.\\ 
\Biggl. + 
{\beta}^{2}\sum\limits_{l=1}^{\infty}\int\limits_{\sqrt{2l|eB|}}^{\infty}dE \sqrt{E^2 - 2l|eB|} \sum\limits_{\pm}\tanh\left[\frac 12\beta\left(E \mp \mu_5\right)\right]{\rm sech}^2\left[\frac 12\beta\left(E \mp \mu_5\right)\right]\Biggr\}
\end{multline}
in the case of $|\mu| \ll \sqrt{2|eB|}, \, |\sqrt{2|eB|} \mp \mu_5|$, and
\begin{multline}\label{3.52}
\left\langle J^{05}\right\rangle_{\beta} = \frac{|eB|}{2\pi^{2}} \mu_5 \Biggl\{1 \Biggr. 
\\ 
\Biggl. + 
{\beta}^{2}\sum\limits_{l=1}^{\infty}\int\limits_{\sqrt{2l|eB|}}^{\infty}dE \sqrt{E^2 - 2l|eB|} \sum\limits_{\pm}\tanh\left[\frac 12\beta\left(E \mp \mu\right)\right]{\rm sech}^2\left[\frac 12\beta\left(E \mp \mu\right)\right]\Biggr\}
\end{multline}
in the case of $|\mu_5| \ll \sqrt{2|eB|}, \, |\sqrt{2|eB|} \mp \mu|$. Thus, contrary to the assertion of \cite{Fuk}, the vector current is not proportional to the axial charge density  in the case of $\sqrt{2|eB|} \gg |\mu_5|$, and this is due to the contribution of the $l \geq 1$ levels. The proportionality emerges even for arbitrary values of $eB$ in the zero-temperature limit only:
\begin{equation}\label{3.53}
\left\langle J^{3}\right\rangle_{\beta}=-{\rm sgn}(eB)\lim\limits_{\beta \to \infty}\left\langle J^{05}\right\rangle_{\beta}, 
\end{equation}
${\rm sgn}(u)=\Theta(u) - \Theta(-u)$ is the sign function.

\section{Explicit violation of chiral symmetry by fermion mass}
\setcounter{equation}{0}

Instead of \eqref{2.1}, let us consider the following integral over the Grassman fields:
\begin{equation}\label{4.1}
{\rm e}^{-\beta \Gamma_{\mu_{5}}[\boldsymbol{\eta}]}=C\int D\psi^{\dagger}D\psi \, 
{\rm exp}\left\{-\int\limits_{0}^{\beta}d\tau \int d^{3}x \, \psi^{\dagger}\left[\partial_{\tau}+H+m\gamma^{0}-\mu_{5}\gamma^{5}+\gamma^{0}\boldsymbol{\gamma}\cdot\boldsymbol{\eta}\left(\boldsymbol{x}\right)\right]\psi\right\}, 
\end{equation}
where $H$ as before is the chirally symmetric Dirac Hamiltonian operator, see \eqref{1.1}, and $m$ is the fermion mass. Performing integration over the Grassman fields, one obtains 
\begin{equation}\label{4.2}
\Gamma_{\mu_{5}}[\boldsymbol{\eta}]=-\frac{1}{\beta}
\int d^{3}x \, {\rm tr}\left\langle\boldsymbol{x}\right|
\ln\cosh\left[\frac{\beta}{2}\sqrt{\left(H-\mu_{5}\gamma^{5}+\gamma^{0}
\boldsymbol{\gamma}\cdot\boldsymbol{\eta}\right)^2 +m^2}\right]\left|
\boldsymbol{x}\right\rangle,
\end{equation}
where relation
\begin{equation}\label{4.3}
\left[H - \mu_{5}\gamma^5, \gamma^{0}\right]_{+}=0
\end{equation}
is taken into account. In view of \eqref{2.12}, one further gets
\begin{equation}\label{4.4}
\Gamma_{\mu_{5}}[\boldsymbol{\eta}]=-\frac{1}{\beta}\sum\limits_{\pm}\int d^{3}x \, {\rm tr}\left\langle \pm,\boldsymbol{x}\right|\ln\cosh\left[\frac{\beta}{2}\sqrt{\left(H_{\pm}\mp \mu_5\mp\boldsymbol{\sigma}\cdot \boldsymbol{\eta} \right)^2+m^2}\right]\left|\boldsymbol{x},\pm \right\rangle.
\end{equation}
In an already standard way by variating \eqref{4.4} and using \eqref{3.3}, one obtains the following expression for the current: 
\begin{equation}\label{4.5}
\left\langle \boldsymbol{J}(\boldsymbol{x})\right\rangle_{\beta,\mu_5}=\frac{1}{2}\int\limits_{-\infty}^{\infty}dE\sum\limits_{\pm}
\boldsymbol{\tau}_{\pm}(\boldsymbol{x}, E)\frac{\pm E-\mu_5}{\sqrt{\left(E\mp\mu_5\right)^2+m^2}}\tanh\left[\frac{\beta}{2}\sqrt{\left(E\mp\mu_5\right)^2+m^2}\right].
\end{equation}
Provided that conditions \eqref{3.27} and \eqref{3.30} are satisfied, one obtainss
\begin{equation}\label{4.6}
\left\langle \boldsymbol{J}(\boldsymbol{x})\right\rangle_{\beta,\mu_5}=\int\limits_{0}^{\infty}dE \, \boldsymbol{\tau}(\boldsymbol{x}, E)\sum\limits_{\pm}
\frac{\pm E-\mu_5}{\sqrt{\left(E\mp\mu_5\right)^2+m^2}}\tanh\left[\frac{\beta}{2}\sqrt{\left(E\mp\mu_5\right)^2+m^2}\right].\end{equation}
In the case of \eqref{3.17} and \eqref{3.19} one obtains \eqref{3.20}. Thus, the chiral magnetic effect in the uniform magnetic field remains unaltered at 
$\mu = 0$ when chiral symmetry is violated by the fermion mass term.

It should be emphasized that the above arguments do not work in the case of the chiral separation effect, since an analogue of \eqref{4.3} (that is with $\mu_{5}\gamma^5$ changed to $\mu$) is absent. However, it is possible to study the chiral symmetry violation perturbatively in the value of mass. 

Let us consider a more general setup with both chemical potentials, $\mu$ and $\mu_5$, being nonvanishing and start from the generating functionals for currents in the following form, cf. \eqref{4.2}, 
\begin{equation}\label{4.7}
\Gamma[\boldsymbol{\eta}]=-\frac{1}{\beta}
\int d^{3}x \, {\rm tr}\left\langle\boldsymbol{x}\right|
\ln\cosh\left[\frac{\beta}{2}\left(H + m \gamma^{0} -\mu  -\mu_{5}\gamma^{5} + \gamma^{0}
\boldsymbol{\gamma}\cdot\boldsymbol{\eta}\right)\right]\left|
\boldsymbol{x}\right\rangle
\end{equation}
and
\begin{equation}\label{4.8}
\Gamma[\boldsymbol{\eta_{5}}]=-\frac{1}{\beta}
\int d^{3}x \, {\rm tr}\left\langle\boldsymbol{x}\right|
\ln\cosh\left[\frac{\beta}{2}\left(H + m \gamma^{0} -\mu -\mu_{5}\gamma^{5} + \gamma^{0}
\boldsymbol{\gamma}\gamma^{5}\cdot\boldsymbol{\eta}_{5}\right)\right]\left|
\boldsymbol{x}\right\rangle.
\end{equation}
In a standard way one can obtain expressions for the vector and axial currents and expand them in powers of the mass value. Note that the odd powers are absent, since they are proportional to traces of off-diagonal matrices, which are zeros. The perturbative corrections to the zero-mass results, $\left\langle \boldsymbol{J}(\boldsymbol{x})\right\rangle_{\beta}$ \eqref{3.25} and $\left\langle \boldsymbol{J}^5(\boldsymbol{x})\right\rangle_{\beta}$ \eqref{3.26}, can be presented as                                                                     
\begin{multline}\label{4.9}
\left\langle \boldsymbol{J}(\boldsymbol{x})\right\rangle_{\beta}^{\left(m^2\right)}=\frac{\beta}{4}\sum\limits_{n=1}^{\infty}\frac{m^{2n}}{(2n)!}\int\limits_{-\infty}^{\infty}dE\sum\limits_{\pm} (\pm)
\boldsymbol{\tau}_{\pm}(\boldsymbol{x}, E)\\
\times\frac{{\partial}^{2n-1}}{{\partial}E^{2n-1}}{\rm sech}^2\left[\frac{\beta}{2}
\left(E-\mu  \mp \mu_5\right)\right]
\end{multline}
and
\begin{multline}\label{4.10}
\left\langle \boldsymbol{J}^5(\boldsymbol{x})\right\rangle_{\beta}^{\left(m^2\right)}=\frac{\beta}{4}\sum\limits_{n=1}^{\infty}\frac{m^{2n}}{(2n)!}\int\limits_{-\infty}^{\infty}dE\sum\limits_{\pm}
\boldsymbol{\tau}_{\pm}(\boldsymbol{x}, E) \\\times\frac{{\partial}^{2n-1}}{{\partial}E^{2n-1}}{\rm sech}^2\left[\frac{\beta}{2}\left(E-\mu  \mp \mu_5\right)\right],
\end{multline}
where the use is made of relation
$$
 \frac{{\partial}^{2n}}{{\partial}u^{2n}}\tanh(c u) = c \, \frac{{\partial}^{2n-1}}{{\partial}u^{2n-1}}{\rm sech}^2(c u).
$$
Provided that conditions \eqref{3.27} and \eqref{3.30} are satisfied, one obtains
\begin{multline}\label{4.11}
\left\langle \boldsymbol{J}(\boldsymbol{x})\right\rangle_{\beta}^{\left(m^2\right)}=\frac{\beta}{4}\sum\limits_{n=1}^{\infty}\frac{m^{2n}}{(2n)!}\int\limits_{0}^{\infty}dE \, \boldsymbol{\tau}(\boldsymbol{x}, E) \\\times \sum\limits_{\pm}(\pm)
\frac{{\partial}^{2n-1}}{{\partial}E^{2n-1}}\left\{{\rm sech}^2\left[\frac{\beta}{2}\left(E - \mu \mp \mu_5\right)\right] + {\rm sech}^2\left[\frac{\beta}{2}\left(E + \mu \mp \mu_5\right)\right]\right\}
\end{multline}
and
\begin{multline}\label{4.12}
\left\langle \boldsymbol{J}^5(\boldsymbol{x})\right\rangle_{\beta}^{\left(m^2\right)}=\frac{\beta}{4}\sum\limits_{n=1}^{\infty}\frac{m^{2n}}{(2n)!}\int\limits_{0}^{\infty}dE \, \boldsymbol{\tau}(\boldsymbol{x}, E) \\
\times \sum\limits_{\pm}(\pm)
\frac{{\partial}^{2n-1}}{{\partial}E^{2n-1}}\left\{{\rm sech}^2\left[\frac{\beta}{2}\left(E \mp \mu - \mu_5\right)\right] + {\rm sech}^2\left[\frac{\beta}{2}\left(E \mp \mu + \mu_5\right)\right]\right\}.
\end{multline}
In the case of \eqref{3.17} and \eqref{3.19} one obtains expressions
\begin{equation}\label{4.13}
\left\langle J^{3}\right\rangle_{\beta}^{\left(m^2\right)} =\frac{e B}{\pi^2}\cosh(\beta \mu)\sinh(\beta \mu_5)\sum\limits_{n=1}^{\infty}\frac{m^{2n}}{(2n)!}\beta^{2n-1}\lim\limits_{E \to \infty} {\rm e}^{-\beta E}
\end{equation}
and
\begin{equation}\label{4.14}
\left\langle J^{35}\right\rangle_{\beta}^{\left(m^2\right)} =\frac{e B}{\pi^2}\sinh(\beta \mu)\cosh(\beta \mu_5)\sum\limits_{n=1}^{\infty}\frac{m^{2n}}{(2n)!}\beta^{2n-1}\lim\limits_{E \to \infty} {\rm e}^{-\beta E},
\end{equation}
that are evidently vanishing. Thus, both the chiral magnetic and chiral separation effects in the uniform magnetic field remain the same as \eqref{3.39} and \eqref{3.40}, if the chiral symmetry violation is treated perturbatively in the value of fermion mass.\footnote{An assertion of the authors of \cite{Met} about the dependence of the chiral separation effect on mass and temperature (see, for instance, formula (36) in \cite{Met}) is not true.}

\section{Persisting currents in the magnetic field with finite flux}

\setcounter{equation}{0}

Let us consider a smooth (without singularities) nonuniform configuration of an external static magnetic field strength. Choosing curvilinear coordinates in such a way that $x^3$ is along the magnetic force lines, while $x^1$ and $x^2$ are orthogonal to them, I assume that the magnetic field strength depends on $x^1$ and $x^2$ only, $B^3 (x^1,x^2)$. Then a surface that is orthogonal at each value of $x^3$ to the magnetic force lines is characterized by the squared length element
\begin{equation}\label{5.1}
{\rm d} s^2 = g_{jj'} {\rm d}x^{j} {\rm d}x^{j'}, \quad j, j' =1,2. 
\end{equation}
The total magnetic flux in the units of $2\pi$ (or $2\pi {\hbar} c$ if the fundamental constants are restored) through the surface,
\begin{equation}\label{5.2}
\Phi = \frac{1}{2\pi} \int d^{2}x \, \sqrt{g} B^3 
\end{equation}
($g = {\rm det}g_{jj'}$), is independent of $x^3$. Let us also define the total curvature (in the units of $2\pi$) of the surface,
\begin{equation}\label{5.3}
\Phi_K = \frac{1}{2\pi} \int d^{2}x \, \sqrt{g} K,
\end{equation}
where $K (x^1,x^2)$ is the Gauss curvature. I assume that the magnetic field strength is concentrated in a central region around $x^1 = x^2 = 0$, decreasing sufficiently fast at large distances  from the center. Then the same can be asserted about the Gauss curvature. Without loss of generality, I also assume that any surface that is orthogonal to the magnetic force lines is noncompact simply-connected.

As to a solution to the Dirac equation in such a magnetic field background, we note that its dependence on $x^3$ is given by a plane wave, ${\rm exp}{({\rm i}k x^3})/\sqrt{2\pi}$. In view of this, operator $H_\pm$ that is defined by \eqref{2.12} can be presented as
\begin{equation}\label{5.4}
H_\pm = H_{\bot, \pm} \mp k\sigma^3, \quad -\infty<k<\infty,
\end{equation}
where
\begin{equation}\label{5.5}
H_{\bot, \pm}=\pm{\rm i}{\boldsymbol{\sigma}}_{\bot}\cdot\left(\boldsymbol{\partial}-{\rm i}e\boldsymbol{A} + \frac{{\rm i}}{2}\boldsymbol{\omega}\right),
\end{equation}
the bundle and spin connections, 
$\boldsymbol{A}$ and $\boldsymbol{\omega}$, as well as $\sigma^1_\bot$ and $\sigma^2_\bot$, depend on $x^1$ and $x^2$ only. Since relation
\begin{equation}\label{5.6}
\left[H_{\bot, \pm},\sigma^3\right]_{+}=0
\end{equation}
holds, a solution to the eigenvalue equation for $H_{\bot, \pm}$,
\begin{equation}\label{5.7}
H_{\bot, \pm} \left\langle \pm, x^1, x^2 \, | \, \varepsilon \right\rangle\ = \varepsilon \left\langle \pm, x^1, x^2 \, | \, \varepsilon \right\rangle\  .
\end{equation}
in the case of either $\varepsilon>0$ or 
$\varepsilon<0$, obeys relation
\begin{equation}\label{5.8}
\sigma^3 \left\langle \pm, x^1, x^2 \, | \,\varepsilon \right\rangle\ = \left\langle \pm, x^1, x^2 \, | \, -\varepsilon \right\rangle\   ,
\end{equation}
So, all eigenfunctions of $H_{\bot, \pm}$ are $\left\langle \pm, x^1, x^2 \, | \,\varepsilon \right\rangle\ $ and $\sigma^3 \left\langle \pm, x^1, x^2 \, | \,\varepsilon \right\rangle\ $. Considering the eigenvalue equation for $H_\pm$,
\begin{equation}\label{5.9}
H_{\pm} \left\langle \pm, \boldsymbol{x} \, | \, E, k \right\rangle\ = E \left\langle \pm, \boldsymbol{x} \, | \, E, k \right\rangle\  ,
\end{equation}
we note that each eigenfunction of $H_{\pm}$ in the case of $\varepsilon \neq 0$ can be presented in general as a superposition of eigenfunctions of $H_{\bot, \pm}$: 
\begin{multline}\label{5.10}
\left\langle \pm, \boldsymbol{x} \, | \, E, k \right\rangle\ = \frac{{\rm e}^{{\rm i}k x^3}}{\sqrt{2\pi}}\left[C_{1, \pm} \left\langle \pm, x^1, x^2 \, | \,\varepsilon \right\rangle\ + C_{2, \pm} \sigma^3 \left\langle \pm, x^1, x^2 \, | \,\varepsilon \right\rangle\ \right], \\ 
E={\rm sgn}(E)\sqrt{\varepsilon^2+k^2}, 
\end{multline}
where
\begin{equation}\label{5.11}
 C_{1, \pm}=\mp{\rm sgn}(Ek)\left(4|E|\right)^{-1/4}\left[|E|-\varepsilon{\rm sgn}(E)\right]^{1/2}
\end{equation}
and
\begin{equation}\label{5.12}
C_{2, \pm}=\left(4|E|\right)^{-1/4}\left[|E|+\varepsilon{\rm sgn}(E)\right]^{1/2}.
\end{equation}
In view of relation
\begin{equation}\label{5.13}
 C_{1, \pm}^2+C_{2, \pm}^2=1,
 \end{equation}
 eigenfunctions of $H_{\pm}$ are properly normalized, as long as eigenfunctions of $H_{\bot, \pm}$ are. Also, one can get 
 \begin{equation}\label{5.14}
\left\langle k, E \, | \, \boldsymbol{x} , \pm \right\rangle\ \sigma^3 \left\langle \pm, \boldsymbol{x} \, | \, E, k \right\rangle\ = \frac {1}{\pi} C_{1, \pm} C_{2, \pm} \left\langle \varepsilon \, | \, x^1, x^2 , \pm \right\rangle\ \left\langle \pm, x^1, x^2 \, | \, \varepsilon \right\rangle\ \sim \mp \frac{k}{E}  .
\end{equation}
 Since \eqref{5.14} vanishes upon summation over the sign of $k$, the contribution of eigenfunctions of $H_\pm$ with 
 $\varepsilon \neq 0$ to 
 \begin{equation}\label{5.15}
\tau_{\pm}^3(\boldsymbol{x}, E)=
{\rm tr}{\sigma}^3\left\langle \left.\pm,\boldsymbol{x}\right|\right.\delta(H_{\pm}-E)\left|\left.\boldsymbol{x},\pm\right.\right\rangle
\end{equation}
vanishes.

Turning now to eigenfunctions of $H_{\pm}$ with $\varepsilon = 0$, one notes that a metric of two-dimensional space can be reduced by coordinate transformations to the conformal form,
\begin{equation}\label{5.16}
{\rm d} s^2 = f(x^1,x^2) \left[\left({\rm d}x^1\right)^2 + \left({\rm d}x^2\right)^2\right]. 
\end{equation}
Zero modes of $H_{\bot, \pm}$ were considered in \cite{Si1}, see also \cite{Si2}. One has
\begin{equation}\label{5.17}
H_{\bot, \pm}=\begin{pmatrix}
0\,&\,L^\dagger_{\pm}\\
L_{\pm}\,&\,0
\end{pmatrix},
\end{equation}
where
\begin{equation}\label{5.18}
L_{\pm} = \pm f^{-3/4} {\rm e}^{-e\phi}\left({\rm i}{\partial}_1 + {\partial}_2 \right)f^{1/4} {\rm e}^{e\phi},
\end{equation}
\begin{equation}\label{5.19}
L^\dagger_{\pm} = \pm f^{-3/4} {\rm e}^{e\phi}\left({\rm i}{\partial}_1 - {\partial}_2 \right)f^{1/4} {\rm e}^{-e\phi},
\end{equation}
and $\phi$ is determined from equation
\begin{equation}\label{5.20}
B^3(x^1,x^2) = \boldsymbol{\Delta} \phi(x^1,x^2),
\end{equation}
$\boldsymbol{\Delta}$ is the Laplace-Beltrami operator. Thus, zero modes consist of two sets that are eigenstates of $\sigma^3$,
\begin{equation}\label{5.21}
\sigma^3 \left\langle x^1, x^2 \, | \, 1, n \right\rangle\ = \left\langle x^1, x^2 \, | \, 1, n \right\rangle\ , \quad \sigma^3 \left\langle x^1, x^2 \, | \, 2, n \right\rangle\ = - \left\langle x^1, x^2 \, | \, 2, n \right\rangle\ .
\end{equation}
The explicit form of square-integrable zero modes is, see \cite{Si1},
\begin{equation}\label{5.22}
\left\langle x^1, x^2 \, | \, 1, n \right\rangle\ = C_{1,n}\,f^{-1/4} {\rm e}^{-e\phi} \left(x^1 - {\rm i}x^2\right)^{n-1}\left(\begin{array}{l} 1\\ 
  0 \end{array} \right), \, n=\overline{1,{\rm integ}_+\left(e\Phi + \frac12\Phi_K\right) - 1}
\end{equation}
and
\begin{equation}\label{5.23}
\left\langle x^1, x^2 \, | \, 2, n \right\rangle\ = C_{2,n}\,f^{-1/4} {\rm e}^{e\phi} \left(x^1 + {\rm i}x^2\right)^{n-1}\left(\begin{array}{l} 0\\ 
  1 \end{array} \right), \, n=\overline{1,-{\rm integ}_-\left(e\Phi - \frac12\Phi_K\right) - 1},
\end{equation}
$n$ is a positive integer, ${\rm integ}_+(u)$ (${\rm integ}_-(u)$) denotes the  closest integer that is more (less) than or equal to $u$; the absolute values of normalization constants are determined from relations   
\begin{equation}\label{5.24}
|C_{1,n}|^2 \int d^{2}x \, f^{1/2} 
{\rm e}^{-2e\phi} r^{2(n-1)} = 1
\end{equation}
and
\begin{equation}\label{5.25}
|C_{2,n}|^2 \int d^{2}x \, f^{1/2} 
{\rm e}^{2e\phi} r^{2(n-1)} = 1,
\end{equation}
$r = \sqrt{(x^1)^2 + (x^2)^2}$. Note that the modes of chirality $\sigma^3 = 1$ are absent for $e\Phi + \frac12\Phi_K \leq 0$, while the modes of chirality $\sigma^3 = - 1$ are absent for $e\Phi - \frac12\Phi_K \geq 0$.\footnote{These results in the case of 
$\Phi_K=0$ were first obtained more than 40 years ago \cite{Aha}, and they are known as the Aharonov-Casher theorem.}

Eigenfunctions of $H_\pm$ with 
$\varepsilon=0$ take form
\begin{equation}\label{5.26}
\left\langle \pm, \boldsymbol{x}; 1 \, | \, \mp k, n \right\rangle\ = \frac{{\rm e}^{{\rm i}k x^3}}{\sqrt{2\pi}}\left\langle x^1, x^2 \, | \, 1, n \right\rangle\ 
\end{equation}
and
\begin{equation}\label{5.27}
\left\langle \pm, \boldsymbol{x}; 2 \, | \, \mp k, n \right\rangle\ = \frac{{\rm e}^{{\rm i}k x^3}}{\sqrt{2\pi}}\left\langle x^1, x^2 \, | \, 2, n \right\rangle\ .
\end{equation}
In view of orthogonality,
\begin{multline}\label{5.28}
\int d^{3}x \, f \left\langle n, \mp k \, | \, 1; \boldsymbol{x}, \pm \right\rangle\  
\left\langle \pm, \boldsymbol{x}; 1 \, | \, \mp k', n' \right\rangle\ =
\int d^{3}x \, f \left\langle n, \mp k \, | \, 2; \boldsymbol{x}, \pm \right\rangle\ 
\left\langle \pm, \boldsymbol{x}; 2 \, | \, \mp k', n' \right\rangle\ \\
= 0, \quad n' \neq n ,
\end{multline}
one obtains the following expressions for the local spectral densities:
\begin{equation}\label{5.29}
\tau_{\pm}^3(\boldsymbol{x}; 1)=
\tau^3(x^1, x^2; 1)=\frac{1}{2\pi} \sum_{n=1}^{{\rm integ}_+\left(e\Phi + \frac12\Phi_K\right) - 1} |C_{1,n}|^2 \, f^{-1/2} {\rm e}^{-2e\phi} r^{2(n-1)}
\end{equation}
and
\begin{equation}\label{5.30}
\tau_{\pm}^3(\boldsymbol{x}; 2)=
\tau^3(x^1, x^2; 2)=-\frac{1}{2\pi} \sum_{n=1}^{-{\rm integ}_-\left(e\Phi - \frac12\Phi_K\right) - 1} |C_{2,n}|^2 \, f^{-1/2} {\rm e}^{2e\phi} r^{2(n-1)}.
\end{equation}
Consequently, with the use of \eqref{4.5}, the chiral magnetic effect takes form
\begin{equation}\label{5.31}
\left\langle J^{3}(x^1, x^2)\right\rangle_{\beta, \mu_{5}}=-2\mu_{5} \left[\tau^3(x^1, x^2; 1) + \tau^3(x^1, x^2; 2)\right].
\end{equation}
Note that the chiral magnetic effect in the magnetic field with finite flux is independent of both $m$ and $\beta$. Note also that, confining to the $m=0$ case and using, instead of \eqref{4.5}, expression \eqref{3.33} that is valid for $\mu \neq  0$, one obtains the same result,
\begin{equation}\label{5.32}
\left\langle J^{3}(x^1, x^2)\right\rangle_{\beta}=-2\mu_{5} \left[\tau^3(x^1, x^2; 1) + \tau^3(x^1, x^2; 2)\right].
\end{equation}

Considering the total vector current, one obtains the following results for it:
\begin{multline}\label{5.33}
\int d^{2}x \, \sqrt{g} \, \left\langle J^{3}(x^1, x^2)\right\rangle_{\beta} \\
=\left\{\begin{array}{l} -\frac{\mu_5}{\pi}\left[{\rm integ}_+\left(e\Phi + \frac12\Phi_K\right) - 1 \right], \quad e\Phi \geq \frac12\Phi_K\\ [6 mm]
 -\frac{\mu_5}{\pi}\left[{\rm integ}_+\left(e\Phi + \frac12\Phi_K\right) + {\rm integ}_-\left(e\Phi - \frac12\Phi_K\right)\right], \, -\frac12\Phi_K < e\Phi < \frac12\Phi_K\\ [6 mm]
- \frac{\mu_5}{\pi}\left[{\rm integ}_-\left(e\Phi - \frac12\Phi_K\right) + 1\right], \quad e\Phi \leq - \frac12\Phi_K\end{array} \right\}
\end{multline}
in the case of $\Phi_K > 0$, and
\begin{multline}\label{5.34}
\int d^{2}x \, \sqrt{g} \, \left\langle J^{3}(x^1, x^2)\right\rangle_{\beta} \\
=\left\{\begin{array}{l} -\frac{\mu_5}{\pi}\left[{\rm integ}_+\left(e\Phi + \frac12\Phi_K\right) - 1\right], \quad e\Phi>-\frac12\Phi_K\\ [6 mm]
 \, \, 0 \, , \quad \frac12\Phi_K \leq e\Phi \leq -\frac12\Phi_K\\ [6 mm]
- \frac{\mu_5}{\pi}\left[{\rm integ}_-\left(e\Phi - \frac12\Phi_K\right) + 1\right], \quad e\Phi<\frac12\Phi_K\end{array} \right\}
\end{multline}
in the case of $\Phi_K \leq 0$; here, it is sensible to distinguish the case of a predominantly convex surface ($\Phi_K > 0$) from that of other surfaces ($\Phi_K \leq 0$).

In a similar way, the chiral separation effect is considered. A result for the axial current is as follows
\begin{equation}\label{5.35}
\left\langle J^{35}(x^1, x^2)\right\rangle_{\beta}=-2\mu \left[\tau^3(x^1, x^2; 1) + \tau^3(x^1, x^2; 2)\right],
\end{equation}
where the local spectral densities are given by expressions \eqref{5.29} and \eqref{5.30}, while results for the total axial current take form
\begin{multline}\label{5.36}
\int d^{2}x \, \sqrt{g} \, \left\langle J^{35}(x^1, x^2)\right\rangle_{\beta} \\
=\left\{\begin{array}{l} -\frac{\mu}{\pi}\left[{\rm integ}_+\left(e\Phi + \frac12\Phi_K\right) - 1 \right], \quad e\Phi \geq \frac12\Phi_K\\ [6 mm]
 -\frac{\mu}{\pi}\left[{\rm integ}_+\left(e\Phi + \frac12\Phi_K\right) + {\rm integ}_-\left(e\Phi - \frac12\Phi_K\right)\right], \, -\frac12\Phi_K < e\Phi < \frac12\Phi_K\\ [6 mm]
- \frac{\mu}{\pi}\left[{\rm integ}_-\left(e\Phi - \frac12\Phi_K\right) + 1\right], \quad e\Phi \leq - \frac12\Phi_K\end{array} \right\}
\end{multline}
in the case of $\Phi_K > 0$, and
\begin{multline}\label{5.37}
\int d^{2}x \, \sqrt{g} \, \left\langle J^{35}(x^1, x^2)\right\rangle_{\beta} \\
=\left\{\begin{array}{l} -\frac{\mu}{\pi}\left[{\rm integ}_+\left(e\Phi + \frac12\Phi_K\right) - 1\right], \quad e\Phi>-\frac12\Phi_K\\ [6 mm]
 \, \, 0 \, , \quad \frac12\Phi_K \leq e\Phi \leq -\frac12\Phi_K\\ [6 mm]
- \frac{\mu}{\pi}\left[{\rm integ}_-\left(e\Phi - \frac12\Phi_K\right) + 1\right], \quad e\Phi<\frac12\Phi_K\end{array} \right\}
\end{multline}
in the case of $\Phi_K \leq 0$. Since the local spectral densities, 
$\tau^3(x^1,x^2;1)$ \eqref{5.29} and $\tau^3(x^1,x^2;2)$ \eqref{5.30}, are independent of energy, results \eqref{5.32}-\eqref{5.37} remain unaltered when chiral symmetry is violated perturbatively in the value of fermion mass.

\section{Conclusions}
\setcounter{equation}{0}

In the present paper, a general formalism for description of chiral effects in thermal equilibrium is elaborated. The key role in my approach is played by the local spectral density, see \eqref{3.4}, that in a straightforward manner is formed from solutions to the Dirac equation. I demonsrate that, in a quantum fermionic system with chiral symmetry \eqref{1.1}, the persisting currents are generated in thermal equilibrium, see \eqref{3.25} and \eqref{3.26}. Under plausible
assumptions \eqref{3.27} and \eqref{3.30} about a behavior of the local spectral density, the axial current is nonvanishing, as long as the chemical potential is, $\mu \neq 0$, while the vector current is nonvanishing, as long as the chiral chemical potential is, $\mu_5 \neq 0$; see \eqref{3.33} and \eqref{3.34}, that remain immutable under simultaneous interchange of 
$\left\langle \boldsymbol{J}^5\right\rangle_{\beta}$ with $\left\langle \boldsymbol{J}\right\rangle_{\beta}$ and of $\mu$ with $\mu_5$. One can conclude that an emergence of the vector current is substantiated on the same grounds as that of the axial current. If chiral symmetry is explicitly violated by the fermion mass term, then a power series of perturbative in the value of mass corrections to the vector or axial current is given by \eqref{4.9} or \eqref{4.10}, in particular, by \eqref{4.11} or \eqref{4.12} under conditions \eqref{3.27} and \eqref{3.30}. A nonperturbative treatment of the chiral symmetry violation at $\mu=0$ is available in view of \eqref{4.3}, and an exact  expression for the resulting vector current is given by \eqref{4.5}, in particular, by \eqref{4.6} under conditions \eqref{3.27} and \eqref{3.30}. 

The elaborated approach is immediately applicable to quantum fermionic systems in the background of a magnetic field. In the case of a static uniform magnetic field, using the complete set of solutions to the Dirac equation, \eqref{3.10}-\eqref{3.15}, one determines the local spectral density, \eqref{3.17} and \eqref{3.19}, and obtain the chiral magnetic effect that is given by the persisting vector current, $\left\langle J^{3}\right\rangle_{\beta}$ \eqref{3.39}, and the chiral separation effect that is given by the persisting axial current, $\left\langle J^{35}\right\rangle_{\beta}$ \eqref{3.40}; here the magnetic field strength is directed along the $x^3$ coordinate axis. Contrary to the assertions in \cite {Yam,Ben}, both effects are substantiated on equal footing (as was previously noted), being independent of temperature, that is a consequence of the energy-independence of the local spectral density. The latter in its turn is due to the fact that only the lowest ($l=0$) Landau level contributes, whereas the contribution of all other ($l \geq 1$) Landau levels is canceled, see \eqref{3.18}. The relevance of exclusively the lowest Landau level, as well as independence of temperature, is sometimes regarded as an evidence for topological nature of the chiral effects. I shall dwell on topological nature later, but now note here that the contribution of the $l \geq 1$ Landau levels to the local spectral density, if non canceling, is necessarily dependent on energy, the latter entailing the temperature dependence for an effect. This is confirmed by my results for the induced temporal components of the currents. The appropriate local spectral density contains a contribution of the $l \geq 1$ Landau levels, see \eqref{3.48}, yielding the temperature-dependent fermion number and axial charge densities, 
$\left\langle {J}^{0}\right\rangle_{\beta}$ \eqref{3.49} and 
$\left\langle {J}^{05}\right\rangle_{\beta}$ \eqref{3.50}; the contribution of the $l \geq 1$ levels dies out in the zero-temperature limit.  

In the case of a static nonuniform magnetic field with finite flux, I find that only zero modes of two-dimensional Hamiltonian operator $H_{\bot, \pm}$, \eqref{5.22} and \eqref{5.23}, contribute, and the local spectral density is given by expressions \eqref{5.29} and \eqref{5.30}. Consequently, the chiral magnetic effect is given by the persisting vector current, $\left\langle J^{3}(x^1, x^2)\right\rangle_{\beta}$ \eqref{5.31}, and the chiral separation effect is given by the persisting axial current, 
$\left\langle J^{35}(x^1, x^2)\right\rangle_{\beta}$ \eqref{5.35}; here curvilinear coordinates are chosen in such a way that $x^3$ is along the magnetic force lines, while $x^1$ and $x^2$ are orthogonal to them. The total vector current is given by expresions \eqref{5.33} and \eqref{5.34}, while the total axial current is given by expresions \eqref{5.36} and \eqref{5.37}. Note that the total currents, in addition to proportionality to the appropriate chemical potentials, depend exclusively on the global characteristics of the background: the total magnetic flux and the total curvature of a surface that is orthogonal to the magnetic force lines. 

The results for the total currents can be presented as
\begin{equation}\label{6.1}
\int d^{2}x \, \sqrt{g} \, \left\langle J^{3}(x^1, x^2)\right\rangle_{\beta}  
= - \frac{\mu_5}{\pi} \, {\rm Index}L_{\pm}
\end{equation}
and
\begin{equation}\label{6.2}
\int d^{2}x \, \sqrt{g} \, \left\langle J^{35}(x^1, x^2)\right\rangle_{\beta} 
= - \frac{\mu}{\pi} \, {\rm Index}L_{\pm},
\end{equation}
where
\begin{equation}\label{6.3}
{\rm Index}L_{\pm} = {\rm DimKer}L_{\pm} - {\rm DimKer}L^\dagger_{\pm},
\end{equation}
${\rm DimKer}L_{\pm}$ (${\rm DimKer}L^\dagger_{\pm}$) is the number of square-integrable zero modes of operator $L_{\pm}$ \eqref{5.18} ($L^\dagger_{\pm}$ \eqref{5.19}). It is relevant to discuss an issue of topology  at this point. 

The index theorem of Atiyah and Singer \cite{Ati1,Ati2,Ati3} establishes a profound interconnection between the analytic properties of differential operators acting on sections of fiber bundles and the topological properties of the fiber bundles themselves. Specifically, the index theorem when applied to quantum field theory allows one to relate the number of normalizable zero modes of the Dirac operator in the background field to the topological characteristics of the background field. The index theorem of Atiyah and Singer is valid for the case of bundle manifolds with closed compact base space, in particular, it takes the following form for a spatial dimensionality equal to 2: 
\begin{equation}\label{6.4}
{\rm Index}L_{\pm} = e \Phi,
\end{equation}
where $\Phi$ is given by \eqref{5.2}.
The right-hand side of \eqref{6.4} is independent of geometry of a surface, i.e., it is the same for a sphere as for any other closed compact surface. Moreover, the value of total magnetic flux through a closed compact surface is quantized, and
\begin{equation}\label{6.5}
e \Phi = N, \quad N = \mathbb{Z}.
\end{equation}
To see this, let us use compactification by stereographic projection of an infinite plane to a sphere of radius $R$,   
\begin{equation}\label{6.6}
X^1 = \frac{2R^2 x^1}{R^2+r^2}, \quad X^2 = \frac{2R^2 x^2}{R^2+r^2}, \quad X^3 = \frac{R(R^2-r^2)}{R^2+r^2}, \quad (X^1)^2+(X^2)^2+(X^3)^2 = R^2. 
\end{equation}
The left-hand side of \eqref{6.5} can be presented through an integral over a circle of infinitely large radius, 
\begin{equation}\label{6.7}
e \Phi = \frac{e}{2\pi} \lim\limits_{r \to \infty}
\ointctrclockwise 
d \boldsymbol{x} \cdot \boldsymbol{A}(\boldsymbol{x}).
\end{equation}
Compactification by means of stereographic projection implies that all points lying on the circle are identified as those corresponding to the south pole of a sphere, $X^1=X^2=0$ and $X^3=-R$. This ensures that 
$\boldsymbol{A}$ asymptotically is a pure gauge \cite{Bel},
\begin{equation}\label{6.8}
\boldsymbol{A}(\boldsymbol{x}) \underset{r \to \infty}{=} \frac{1}{{\rm i} e} g^{-N} \, \boldsymbol{\partial} \,  g^{N}. 
\end{equation}
The unambiguity of an element of gauge transformations results in
\begin{equation}\label{6.9}
 g^{N} = {\rm e}^{{\rm i} N \vartheta}, \quad N = \mathbb{Z}
\end{equation}
($\vartheta={\rm arctan}(x^2/x^1)$), and the latter yields \eqref{6.5}. Thus, configurations of a magnetic field through a closed compact surface are divided into sets of topologically equivalent configurations, with each set characterized by its value of winding number $N$. 

Returning to the case of manifolds with noncompact base space, we recall that the global bundle characteristics, see \eqref{5.2}, as well as the global characteristics of a base space, see \eqref{5.3}, can take arbitrary values. Nevertheless, an analogue of the index theorem exists for a spatial dimensionality equal to 2: it is given in \cite{Si1} for simply-connected noncompact surfaces, and in \cite{Mi} for more general (topologically finite) noncompact surfaces. The index in the former case is equal to the factor that is after $-\mu_5/\pi$ in \eqref{5.33} and \eqref{5.34}, or after $-\mu/\pi$ in \eqref{5.36} and \eqref{5.37}. One can speak about topological nature of chiral currents in magnetic field as long as they are related to the index. However, as was just demonstrated, there is a crucial distinction of the analogue of the index theorem for the noncompact case from the true index theorem for the closed compact case: although only global characteristic are involved in both cases, any topological specification (division into classes differing by winding number) is lacking in the noncompact case. Moreover, even with restriction to the particular case of 
\begin{equation}\label{6.10}
\Phi_K = 0, \quad e \Phi = N, 
\end{equation} 
the ``index theorem'' takes the form
\begin{eqnarray}\label{6.11}
{\rm Index}L_{\pm}=
\left\{\begin{array}{l} N - {\rm sgn}(N), \quad  N \neq 0,\\ 
 \quad \quad  0, \quad \quad N=0, \end{array} 
\right.
\end{eqnarray}
which differs from that for the closed compact case, see \eqref{6.4} and \eqref{6.5}:
\begin{equation}\label{6.12}
{\rm Index}L_{\pm} = N.
\end{equation} 
Therefore, an appeal towards the index theorem in the arguments to substantiate theoretically chiral effects in magnetic field can be regarded as no more than heuristic.

Finally, it should be emphasized once more that the local spectral density most adequately encodes the basics of chiral effects. A point is the energy-independence of the local spectral density, that entails the temperature-independence of the chiral effects. On the one hand, it is owing to the exclusive relevance of the lowest Landau level in the uniform magnetic field, or, appropriately, of the zero modes in the nonuniform magnetic field with finite flux, the latter providing a link to the apparent topological aspect of the chiral effects. On the other hand, due to the energy-independence of the local spectral density, the chiral effects in magnetic field remain unaltered if chiral symmetry is explicitly violated by the fermion mass term. This is proven in the framework of the perturbation theory expansion in the value of mass. In view of \eqref{4.3}, the same is also proven at $\mu = 0$ in the framework of the nonperturbative treatment for the chiral magnetic effect.

\section*{Acknowledgments}

I am thankful to Daniel Boer and Aitzol Garcia Etxarri for stimulating discussions and valuable remarks. I acknowledge support from IKUR Strategy under the collaboration agreement between Ikerbasque Foundation and DIPC on behalf of the Department of Education of the Basque Government, Programa de ayudas de apoyo a los agentes de la Red Vasca de Ciencia, tecnología e Innovación acreditados en la categoría de Centros de Investigación Básica y de excelencia (Programa BERC) Departamento de Universidades e Investigación del Gobierno Vasco and the Centros Severo Ochoa AEI/CEX2018-000867-S from the Spanish Ministerio de Ciencia e Innovación. My work was also supported by the National Academy of Sciences of Ukraine under Project No. 0122U000886.

\end{document}